\documentclass[a4paper]{article}

\usepackage{INTERSPEECH2021}
\usepackage{float}
\usepackage{threeparttable}

\title{Accented Speech Recognition Inspired by Human Perception}
\name{Xiangyun Chu$^1$, Elizabeth Combs$^1$, Amber Wang$^1$, Michael Picheny$^2$}

\address{
  $^1$Center for Data Science, New York University \\
  $^2$Courant Computer Science and Center for Data Science, New York University}
\email{\{xc1511, eac721, yw2115, map22\}@nyu.edu}

\begin{document}

\maketitle
\begin{abstract}
    While improvements have been made in automatic speech recognition performance over the last several years, machines continue to have significantly lower performance on accented speech than humans. In addition, the most significant improvements on accented speech primarily arise by overwhelming the problem with hundreds or even thousands of hours of data. Humans typically require much less data to adapt to a new accent. This paper explores methods that are inspired by human perception to evaluate possible performance improvements for recognition of accented speech, with a specific focus on recognizing speech with a novel accent relative to that of the training data. Our experiments are run on small, accessible datasets that are available to the research community. We explore four methodologies: pre-exposure to multiple accents, grapheme and phoneme-based pronunciations, dropout (to improve generalization to a novel accent), and the identification of the layers in the neural network that can specifically be associated with accent modeling. Our results indicate that methods based on human perception are promising in reducing WER and understanding how accented speech is modeled in neural networks for novel accents. 
\end{abstract}

\section{Introduction}




Although humans can quickly adapt to different accents, recognition of accented speech remains a daunting task for today's speech recognition systems. Unlike noise, an accent is an intrinsic, speaker-dependent quality of speech, and humans are capable of understanding a novel accent within one minute of exposure \cite{clarke}. However, machines require hundreds or even thousands of hours of speech data to get good performance \cite{accent-speech5, grapheme-google}. This paper seeks to explore techniques inspired by human learning that go beyond merely gathering massive amounts of additional data to improve word error rate (WER) for accented speech recognition. The four methodologies we explore are all inspired by how humans learn to adapt to accented speech. 

First, humans with pre-exposure to multiple accents can often perceive new, accented pronunciations more easily compared to individuals with limited or no accent exposure \cite{clarke}. Therefore, we conduct experiments to compare a model primarily trained on native speech against a model with substantial pre-exposure to accented speech for the recognition of speech from a novel accent.  
Second, non-native speakers of English often pronounce words as they are spelled, at least until they are more familiar with the complexities of English pronunciation \cite{grapheme-human1,grapheme-human2}. Thus, the second model improvement that we investigate is acoustic modeling using both grapheme and phoneme lexicons during pretraining and adaptation. 


Third, after exposure to a prior set of accents, humans are able to generalize to a new accent more robustly \cite{clarke}. Our experiments examine the regularization effect of this prior knowledge of accents to help our models generalize to a novel accent. To further enable this generalization in neural networks, one common regularization technique is to use dropout \cite{hinton-dropout}. We explore adjusting the amount of dropout during pretraining and adaptation. 
Fourth, while we know that accented speech is perceived differently than noisy speech \cite{clarke}, it is less well understood when and how the accent is processed in both perceptual and machine learning. Therefore, we explore updating different layers of the neural network to assess where accents are most represented.  


\subsection{Related Work}

There have been many attempts to improve the recognition of accented speech, with varying degrees of success \cite{accent-speech1,accent-speech2,accent-speech3, accent-speech9,accent-speech10}. Some promising approaches include unsupervised adaptation \cite{unsupervised1, accent-speech8}, multitask learning with accent embeddings \cite{accent-speech6,accent-speech7}, and domain adversarial training \cite{accent-speech5,accent-speech4}.
While most approaches have delivered results, they either use massive amounts of accent data (e.g., 23K hours \cite{accent-speech5}), rely on corpora that are not publicly available \cite{accent-speech5,grapheme-google}, or use increasingly complex models \cite{accent-speech9,accent-speech6,accent-speech4} that do not shed light on how humans adapt so quickly to new accents. 
Generally speaking, humans are able to very quickly adapt to a wide variety of speech \cite{clarke}. Previous research into human perception in native and non-native discourse suggests that people benefit from contextual familiarity of topic, accent, or speaker flexibility \cite{familiarity, berk}. 

The inconsistent relationship between spelling and pronunciation of English creates additional challenges to non-native speakers of English
\cite{grapheme-human1,grapheme-human2}. They will often pronounce words according to their orthography rather than their ``proper" pronunciations. Previous work shows that grapheme-based models may be a better alternative to traditional phoneme-based models for non-native speakers, but the approach typically requires a large amount of data (usually 3K to 12K hours) and more complex neural models \cite{grapheme-google,grapheme-paper1,grapheme-FB}.

Generally speaking, deeper neural networks tend to perform better than shallower networks on automatic speech recognition (ASR) \cite{layer1, layers-tsne}. The lower layers of the deep acoustic models are thought to extract low-level spectral features and upper layers to extract high-level information (e.g., phonemes and words) \cite{layers-tsne, layers-human}. 
For speaker adaptation, models benefit even if only the input or the output layers are updated \cite{layer-first}.
Previous work on accent adaptation has also leveraged the assumption that most benefit accrues by focusing adaptation on the top layers of the network \cite{layer2}. However, few studies have examined in detail which layers are most effective in accent adaptation.
Therefore, our experiments investigate thoroughly individual and grouped layers of the deep network to identify which layers are most associated with accent for the best model improvement.

\section{Methodology}
 
\subsection{Data}
For our novel accented corpus, we utilized the challenging \textit{MALACH} corpus \cite{malach}, a 160-hour set of interviews from 682 Holocaust witnesses, distributed by the LDC. The accents primarily derive from Central and Eastern Europe. MALACH was sampled at different sizes (10, 20, 40, 160 hours) for adaptation. There is also a development set (1.5 hours) that is used for evaluation. 

The corpora used to develop \textit{source} models (the acoustic models we pretrained for recognition of MALACH) are also all available to the research community. We used two other datasets: one primarily containing accented speech, and one primarily containing native speech. We built a 70-hour \textit{Accent-Combined} dataset sampled from three accented corpora: the Mozilla Common Voice corpus \cite{commonvoice}, the Singapore English National Speech Corpus \cite{singapore}, and the L2-Arctic Corpus \cite{arctic} according to the specifications in Table \ref{table:data}. The Common Voice corpus consists of both native and non-native speech of multiple languages. We filtered the data to include only the English language component of the corpus; this contains 15 accents in addition to native US English. The Singapore English corpus includes three parts with each containing 1000 hours of either read or conversational speech data. 
We also processed 150 hours of prompted read speech from Part2 - Channel 0 of the corpus.
The L2-Arctic corpus includes recordings from 24 non-native English speakers with 6 different accents reading the CMU’s ARCTIC prompts. Therefore, the model pretrained on the Accent-Combined dataset simulates a listener with exposure to many accents, none of them matching the accents in the MALACH corpus. In terms of a corpus primarily consisting of native speech, we utilized \textit{LibriSpeech} \cite{librispeech}, a 1000-hour read speech corpus based on audio books (referred to here as \textit{Full-LibriSpeech}), and the supplied 100-hour  ``clean" subset, referred to here as \textit{Mid-LibriSpeech}. Thus, the model pretrained on the LibriSpeech dataset, especially the Mid-LibriSpeech subset, simulates a listener with only native speech exposure.





\begin{table}[H]
  \caption{Accent-Combined Corpus Creation}
  \label{table:data}
  \centering
  \begin{tabular}{ccccc}
    \toprule
    Corpus & Train & Accent \\
    \midrule
    Common & 22.8h & Singapore, Australia, Bermuda, \\Voice&&Malaysia,
    England, Hong Kong, \\&&Africa, Ireland, New Zealand, \\&&Canada, Philippines, Scotland,\\&&India, South Atlantic, Wales\\
    \hline
    Singapore & 24.2h & Singapore\\
    \hline
    L2-Arctic & 22.8h & Arabic, Chinese, Hindi,\\&& Korean, Spanish, Vietnamese \\

    \midrule
    Total & 69.9h   \\
    \bottomrule
  \end{tabular}
\end{table}

\subsection{Models}

We chose to use the Kaldi toolkit \cite{kaldi} for building speech recognition models because it offers a hybrid model. In a hybrid model, the acoustic model is separated from the language model and can be adapted separately.  
Our experiments  employed similar Kaldi chain-model recipe set-ups \cite{kaldi-nnet3} to train the three source models (MALACH, Accent-Combined and LibriSpeech). The MALACH source model used a 7-layer TDNN \cite{peddinti} as in Kaldi's MALACH recipe. 
For the Accent-Combined source model, we 
also utilized the same 7-layer TDNN structure as in the MALACH recipe. The LibriSpeech 
source model used a 17-layer TDNN using Kaldi's LibriSpeech recipe.

All adaptation experiments were performed by simply fine-tuning a pretrained source model. 
During adaptation, our experiments employed two different learning rate factors as suggested in Kaldi: a primary and a secondary learning rate. The primary learning rate factor (pLR) controls the opportunity for weight update within the overall architecture of the neural network. The secondary learning rate factor (sLR) controls the weight updates for only the last layer of the network, which is often thought to be the most task-dependent layer. The default values in Kaldi are set to pLR=0.25 and sLR=1.0, thus updating the lower network layers more slowly than the output layer. 

\section{Experiments}


\subsection{Effect of Pre-Exposure to Multiple Accents}

 
 To test the hypothesis that pre-exposure to multiple accents enables ASR to achieve a higher degree of robustness on novel accented speech, we compared the WER of the model trained on the Accent-Combined dataset to the WER of the native speech models trained on LibriSpeech. The target data was MALACH, which contains accented speech not existing or prevalent in the two source datasets.
 
 Since our experiments focused on improving acoustic model performance, we used the fully trained language model that was built for MALACH in all of our experiments. This language model incorporated a MALACH-specific vocabulary. Experiments using this language model without further adaptation in the acoustic model are referred to as \textit{Unadapt} across experiments. We ran additional experiments to assess the benefit of adapting the pretrained acoustic models with MALACH training data. These models are referred to as \textit{Adapt} (40 hours) and \textit{Full-Adapt} (160 hours). A network fully trained using MALACH data served as the baseline performance level that we benchmark against in our experiments. The hyperparameters we tuned include learning rates and number of adaptation epochs.
\begin{table}[H]
  \centering
  \addtolength{\leftskip}{-4cm}
  \addtolength{\rightskip}{-4cm}
  
  \caption{Experimental Results WER (\%) using Unadapted and Adapted Source Models}
  \label{table:experiment1}
  \begin{tabular}{cccc}
    \toprule
    Source Model & Unadapt  & Adapt & Full-Adapt  \\
    WER (\%) & 0h & 40h & 160h \\
    \midrule
     MALACH (160h) & 23.9 & - & - \\
    Accent-Combined (70h) & 39.5 & 33.3 & 30.0  \\
    Mid-LibriSpeech (100h) & 47.2 & 32.4 & 29.1  \\
    Full-LibriSpeech (1000h) &  30.8 & 29.7 & 27.4  \\
    \bottomrule
  \end{tabular}
\end{table}
Table \ref{table:experiment1} shows the results from experiments with and without adaptation of
the different source models. Not completely unexpectedly, the fully-trained MALACH model achieved the best performance with a 23.9\% WER. Without adaptation, using the Mid-LibriSpeech model consisting of 100 hours of primarily native speech, the WER was 47.2\%. Using the Accent-Combined source model with 70 hours of training data, the WER was 39.5\%. This result is consistent with human performance, which suggests that prior accent exposure improves robustness to novel accents. However, even with as little as 40 hours of accent-specific adaptation data, very large improvements (-6.2\% and -14.8\% WER) resulted for both the Accent-Combined and the Mid-LibriSpeech source models. In addition, the much larger source model consisting of the full 1000-hour LibriSpeech corpus containing a broad variety of speech seemed to produce the best results by far apart from the fully-trained MALACH model.

\subsection{Grapheme-Based Models}

We compared performance of phoneme-based acoustic models and grapheme-based models on accented data by modifying the conventional phoneme-based lexicon file in the Kaldi recipe to include letter spellings. We re-trained the Accent-Combined model using only a grapheme lexicon and also jointly using a combined phoneme and grapheme lexicon. The MALACH lexicon was also modified in a similar fashion. 

Table \ref{table:experiment_graph} shows MALACH results using the Accent-Combined source model. The grapheme-only model performed worse than the phoneme model for all adaptation conditions, but the combined lexicon performed somewhat better for the unadapted condition. It appears as if any sort of accent-dependent adaptation compensated for any issues with the lexicon, and perhaps somewhat more so for the phoneme-based lexicon than the grapheme-only lexicon. At least for the relatively small source model corresponding to the Accent-Combined corpus, no particular advantage was seen for graphemes.


\begin{table}[H]
  \centering
  \addtolength{\leftskip}{-2cm}
  \addtolength{\rightskip}{-2cm}
  \caption{Comparison of Phoneme- and Grapheme-based Accent-Combined Models on MALACH. Adaptation methods include unadapted acoustic model with 0h MALACH data and fully adapted acoustic model with 160h MALACH data. Fully adapted model trains for 2 and 4 epochs.}
  \label{table:experiment_graph}
  \begin{tabular}{lccc}
    \toprule
    Adapt. & Phoneme & Grapheme & Grapheme \\WER (\%)& & & + Phoneme \\
    \midrule
     Unadapt & 39.5 & 40.1 & 38.0  \\
     Full-Adapt (2 Eps) & 30.8 & 31.5 & 30.9 \\
     Full-Adapt (4 Eps) & 30.0 & 31.3 & 30.1 \\
    \bottomrule
  \end{tabular}
\end{table}



    

\subsection{Adjusting Dropout}



We examined the hypothesis that pre-exposure to accented speech is equivalent to a form of regularization of the source acoustic model. Regularization is known to improve the generalization of deep learning models to novel test data \cite{Goodfellow-et-al-2016}. Therefore, we applied dropout, a simple way to regularize a model \cite{hinton-dropout}. Kaldi provides a variety of ways of incorporating dropout into the training process. 

Specifically,  we adopted the time-varying dropout schedule `0,0@0.20,0.5@0.50,0' provided as part of the LibriSpeech training recipe. This schedule specifies a dynamic dropout setting for one training epoch in which the probability that a random unit within the network would be dropped gradually increases from 0 to 0.5 after 20\% of the data are seen during the first half of the training epoch and then gradually decreases back to 0 during the second half of the training epoch \cite{Cheng2017}. This dropout schedule was applied when pretraining the source Accent-Combined model; we  refer to this as \textit{Source Dropout}. In order to boost adaptation robustness, the same dropout setting was also applied when updating the pretrained acoustic model on MALACH training data; we refer to this as \textit{Adaptation Dropout}. Moreover, we conducted hyperparameter tuning to search for the optimal set of learning rates and a suitable adaptation training time under the dropout setting.

Table \ref{table:experiment3} shows that adding source dropout decreases the unadapted WER performance on MALACH by 1.6\% (from 39.5\% to 37.9\%). When we updated the acoustic model using 40 hours and 160 hours of MALACH data, the addition of the dropout schedule in both the pretraining and adaptation process decreases the WER by 1.7\% and 0.4\% respectively under default hyperparameter settings. 
We also observed that with dropout added, we could further lower the WER to 29.8\% by increasing the number of adaptation epochs to 8 and lowering the secondary learning rate to 0.5. However, we could not do a full grid search to optimize these parameters for every possible hyperparameter, so we only report on a representative set. 

As a follow-up experiment, we also tried dropout for the source model trained on ``clean", native speech (Mid-LibriSpeech). The benefit of dropout was less obvious. After adding the same dropout schedule when pretraining and updating the source Mid-LibriSpeech model, we only observed a drop of 0.03\% in unadapted WER performance. No improvement was observed after we adapted the source model using MALACH data.

\centerline{}
\begin{table}[H]
  \centering
  \addtolength{\leftskip}{-2cm}
  \addtolength{\rightskip}{-2cm}
  
  \caption{Comparison of Pretrained Accent-Combined Models on MALACH Before and After Dropout. Hyperparameters tuned for different configurations are epochs (ep), primary learning rate (pLR), and secondary learning rate (sLR). The table below uses configuration ep=2, pLR=0.25, sLR=1.0. }
  \label{table:experiment3}
  \begin{tabular}{ccccccc}
    \toprule
    Adapt.  & Source  & Adapt. &  
    WER  \\
    Type & Dropout  & Dropout & 
     (\%) \\
    \midrule
    Unadapt &  N & N &  
    39.5  \\
    Adapt (40h)& N & N & 
    33.3 \\
    Full-Adapt (160h)& N & N & 
    30.8  \\
    \midrule
    Unadapt & Y  & N & 
    37.9  \\
    Adapt (40h)& Y  & Y & 
    31.6  \\
    Full-Adapt (160h)& Y  & Y & 
    30.4  \\
    \bottomrule
  \end{tabular}
\end{table}

\subsection{Updating Neural Network Layers} 
To assess how accent knowledge is captured in the neural network model, we studied which part of the network is most effective in accent adaptation. While lower layers are thought to capture low-level speech features and upper-layers to capture high-level features \cite{layers-tsne}, accent may correspond to a more intermediate aspect of the network. Our methodology was to freeze and unfreeze different layers in different combinations during the adaptation process. These experiments focused on the Full-LibriSpeech source model because of its larger number of layers relative to the Accent-Combined source model. 

\begin{figure}[h]
    \centering
    \includegraphics[width=8cm,height=4cm]{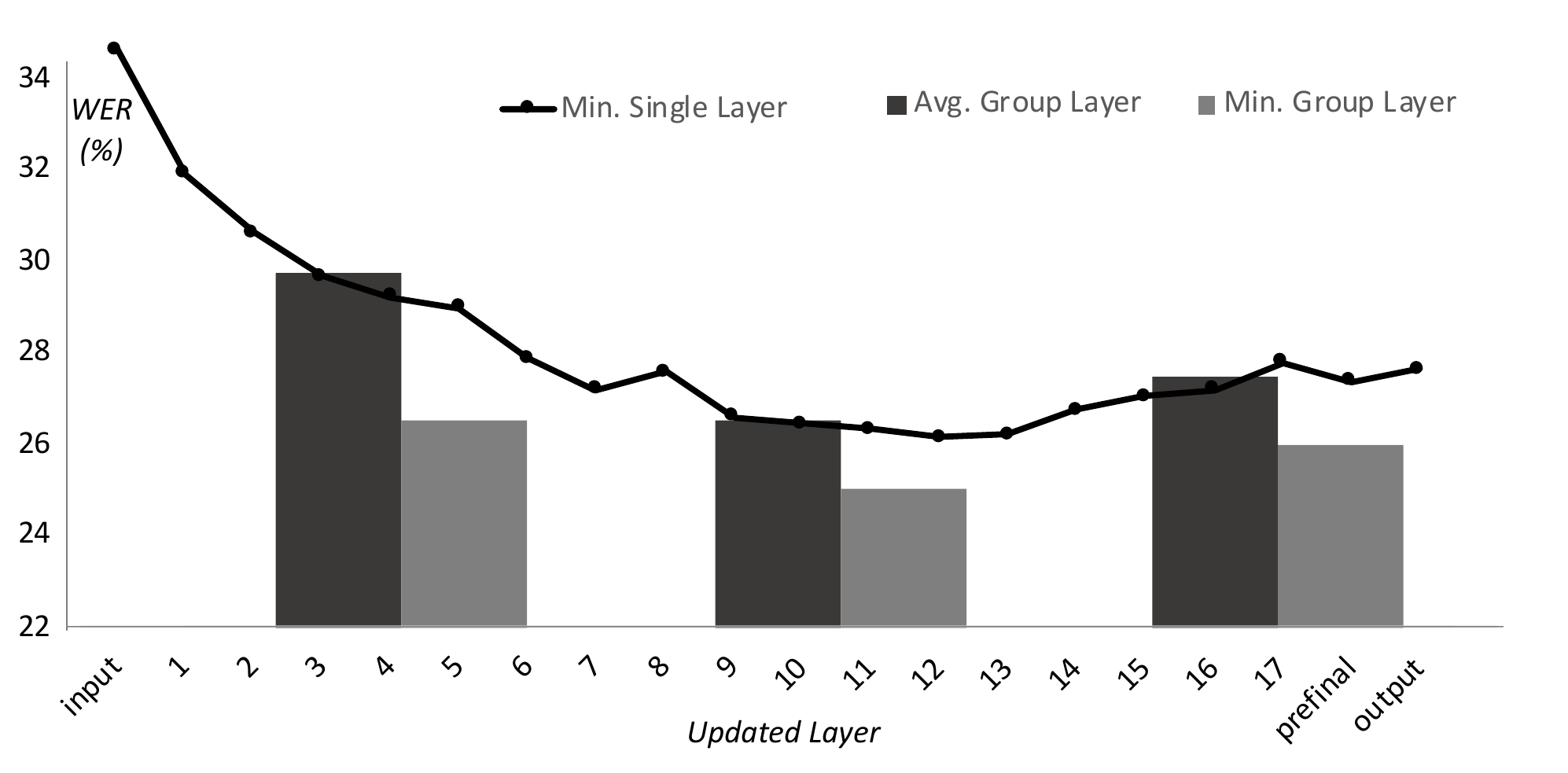}
    \caption{Minimum Experimental WER (\%) Achieved with Isolated Single and Grouped Layer Updates using ep=10, pLR=0.0, sLR=0.125.}
    \label{fig:layer_updates}
\end{figure}

First, we investigated adapting individual layers. We updated one layer at a time with different learning rates (sLR=\{0.125, 0.25, 1.0\}) while freezing all other layers (pLR=0.0). The results are shown by the dark line in Figure \ref{fig:layer_updates}. WERs ranged from 34.6\% to 26.2\% using this methodology (ep=10, pLR=0.0, sLR=0.125). These experiments indicated that the lower layers of the network do not impact WER as much as the middle and upper layers when freezing the other layers completely. Updating the middle layers (layers 9-15) had the most impact with a WER 26.2\% at the center of the network (layer 12). Longer training time and slower learning rates (sLR$<=$0.25) further improved performance. The best run with single layer adaptation (layer 13) achieved a WER of 26.1\% (-1.3\% compared to the default adaptation benchmark of 27.4\% in Table \ref{table:experiment1}), at ep=20 and sLR=0.25. 


Then, we examined grouped layers. Based on the above experiments, we identified three groups of layers within the network: lower (layers 1-8), middle (layers 9-15), and upper (layers 16-output) groups. Similar to updating one layer at a time, results indicated that the lower layer group may be less associated with accent and the middle layer group improved WER the most.
In Figure \ref{fig:layer_updates}, the bar chart shows that updating grouped layers seemed to further improve the WER compared to updating a single layer in isolation.
The best run with grouped layers achieved a WER of 25.1\% (-2.3\% compared to default benchmark), updating the middle layer group with ep=10 and sLR=0.125.


\section{Discussion}
As shown in Table \ref{table:experiment1} above, in the absence of adaptation data, pre-exposure to multiple accents demonstrates clear improvements (-7.7\%) in recognizing a novel accent compared to exposure to ``clean", native speech (Mid-LibriSpeech) for roughly the same amount of data. However, Full-LibriSpeech, performs extremely well with no adaptation, even taking into account that it has 10x the amount of data as Mid-LibriSpeech. One possible explanation is that Full-LibriSpeech is not only larger, but much more diverse than Mid-LibriSpeech in terms of including both noisy and accented speech, effectively also providing a fair amount of accent pre-exposure.  

\begin{table}[H]
  \caption{Comparison of Experimental Results on Accent-Combined Source Models}
  \label{table:experiment_all}
  \centering
  \begin{tabular}{lccc}
    \toprule
    Method &  Unadapt & Full-adapt & Best Config. \\ & WER (\%) & WER (\%)&(Ep, pLR, sLR) \\
    \midrule
    Benchmark & 39.5 & 30.0 & 4, 0.25, 1.0\\
    G + P & 38.0 & 30.1 & 4, 0.25, 1.0\\
    Dropout & 37.9 & 29.8  & 8, 0.25, 0.5\\
    \bottomrule
  \end{tabular}
\end{table}

Table \ref{table:experiment_all} summarizes the improvements we obtained from a combined grapheme-phoneme lexicon and the addition of dropout to the Accent-Combined source model. Both of these techniques provide some degree of robustness in the unadapted case, but are washed out when significant adaptation data is available. We also tried various configurations of the grapheme-based model on LibriSpeech, including adding word boundaries \cite{grapheme-FB}, but none of them made much difference when it came to recognizing MALACH data. Compared to the regularization effect of prior knowledge in accents, the benefit of dropout is less significant. This suggests that accent pre-exposure provides richer information and more structured regularization than dropout, which is uniform and unstructured in its nature.

 While previous studies assumed that accent is best modeled by the top layers of the network \cite{layer2}, our experiments (Figure \ref{fig:layer_updates}) obtained the best results by adapting middle layers of the network. This suggests that accent may be an intermediary quality of speech rather than low-level or high-level information within the network. Using middle-layer feature-processing for accented ASR mirrors human speech perception, which also uses a hierarchical process of  recognition to convert sound to understanding \cite{layers-human}. The fact that different layers of an acoustic model may model different processes may also be relevant to other types of speech adaptation mechanisms, such as speech style and domain adaptation. 
 
 Although in these experiments improvements are generally modest, the results are consistent with human ability to perform well on novel accented speech with little or no adaptation data. This suggests that multiple parallel mechanisms may be in place for human accent perception; more study of which may provide additional insights into improving speech recognition systems.

There are additional steps we would like to take in future work. First, we want to create a true test set for the MALACH data, which could serve as a final benchmark for improvements without the conflating issue of hyperparameter tuning. Second, we would like to repeat these experiments on other accented speech corpora to provide additional support for the findings reported in this paper. Third, we would like to explore methodologies for rapid accent adaptation, such as those based on accent embeddings, to try to allow us to significantly reduce the amount of adaptation data needed for novel accents; 40 hours of adaptation data is still a significant amount of data and seems to be more than required by humans.


\section{Acknowledgements}
This work was supported by the New York University Center for Data Science and originally proposed by Professor Michael Picheny. Experiments
were carried out using and with support from New York University's Courant Institute of Mathematical Sciences (CIMS). 

\bibliographystyle{IEEEtran}
\bibliography{mybib}

\end{document}